\documentclass{XrU2005}

\usepackage{graphicx}
\usepackage{xspace}
\usepackage{amsmath}
\usepackage{natbib}

\graphicspath{{/home/kreyken/mytex/figures/}}

\newcommand{\ca}{\ensuremath{\sim}}
\newcommand{\xte}{{\sl RXTE}\xspace}
\newcommand{\kev}{ke\kern -0.09em V\xspace}
\newcommand{\integral}{{\sl INTEGRAL}\xspace}
\newcommand{\V}{V\,0332\ensuremath{+}53\xspace}

\title{Observation of \V over the 2004/2005 outburst with \integral}
\author[1,2]{I. Kreykenbohm}
\author[2]{N. Mowlavi}
\author[3]{K. Pottschmidt}
\author[4]{J. Wilms}
\author[2,5]{S. E. Shaw}
\author[3]{R. E. Rothschild}
\author[2]{N. Produit}
\author[6]{W.~Coburn}
\author[7]{P. Kretschmar}
\author[1]{A. Santangelo}
\author[1]{R. Staubert}

\affil[1]{Institut f\"ur Astronomie und Astrophysik --
  Astronomie, University of T\"ubingen, 72076 T\"ubingen, Germany}
\affil[2]{INTEGRAL Science Data Centre (ISDC), 1290 Versoix,
  Switzerland} 
\affil[3]{CASS, University of California, San Diego, La Jolla, CA
  92093, U.S.A.}
\affil[4]{Department of Physics, University of Warwick, Coventry, CV4 7AL, UK}
\affil[5]{School of Physics and Astronomy, University of Southampton, SO17 1BJ, UK}
\affil[6]{Space Sciences Laboratory, University of California,
Berkeley, Berkeley, CA, 94702-7450, U.S.A.}
\affil[7]{European Space Astronomy Centre (ESAC), European Space
  Agency, 28080 Madrid, Spain}

\begin{document}


\maketitle

\begin{abstract}
  We present the spectral and temporal analysis of the 2004/2005
  outburst of the transient X-ray pulsar \V as observed with
  \integral.  After the discovery of the third cyclotron line in phase
  averaged spectra \citep{kreykenbohm05a,pottschmidt05a},
  detailed pulse phase spectroscopy revealed remarkably little
  variability of the cyclotron lines through the 4.4\,s X-ray pulse
  \citep{pottschmidt05a}. During the decline of the outburst, the flux
  was observed to decay exponentially until 2005 Feb 10 and linearly
  thereafter. The spectrum was found to become harder with time, while
  the folding energy remained constant.  The energy of the fundamental
  cyclotron line increased with time from 26.5\,\kev in the \xte
  observation up to 29.5\,\kev in the last \integral one indicating
  that the emission region is moving closer to the surface of the
  neutron star. For a detailed analysis, see \citet{mowlavi05a}.
\end{abstract}

\parskip0.5\baselineskip

\section{Introduction: \V}

  The recurring transient X-ray pulsar \V was discovered in 1983 in
  \textsl{Tenma} data \citep{tanaka83a}. Subsequently, a larger
  outburst was found to have occurred in the summer of 1973 when
  analyzing \textsl{Vela 5B} data \citep{terrell84a}. The analysis
  revealed a 4.4\,s pulse period and an indication for a 34.25\,d
  orbital period \citep{stella85a}. The optical counterpart is the
  O8--9 star BQ~Cam \citep{negueruela99a}.
  
  Analysis of the \textsl{Tenma} data revealed a spectral shape
  similar to that seen in other accreting X-ray pulsars with a flat
  power law, an exponential cutoff, and a cyclotron resonant
  scattering feature (CRSF) at an energy of $\sim$28\,keV. In 1989
  September the source experienced another outburst, this time
  observed by \textsl{Ginga} \citep{makino89a}.  With the energy range
  of the Large Area Counters adjusted to cover the 2--60\,keV range,
  CRSFs were detected at 28.5 and 53\,keV.

  Most recently, V\,0332+53 went into outburst in 2004 November and
  was seen by the \textsl{RXTE}/All Sky Monitor (ASM) to reach an
  intensity of $\sim$1\,Crab in the 1.5--12\,keV band
  \citep{remillard04a}. A long series of observations with
  \textsl{RXTE} and \integral were made throughout the outburst.

\begin{figure}
\centerline{\includegraphics[width=0.9\columnwidth]{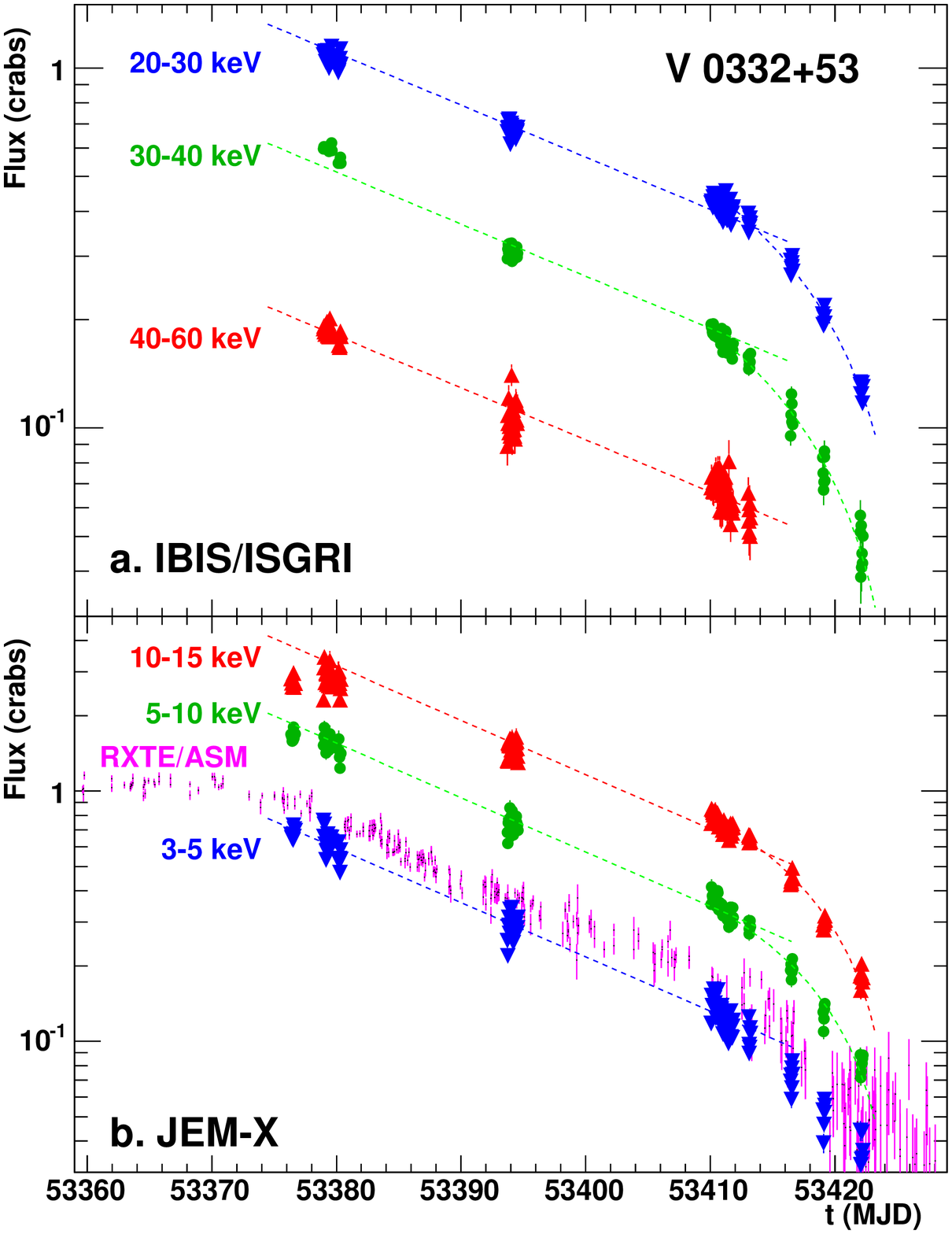}}
\caption{Flux evolution for \V as observed by the \integral
  instruments ISGRI and JEM-X \citep{mowlavi05a}. }
\label{fig:flux}
\end{figure}

\section{Flux evolution}

During the decline phase, the observed fluxes first decay
exponentially up to MJD\,53412, followed by a linear decrease (see
Fig.~\ref{fig:flux}). The decay timescales are different at lower and
higher energies: while a decay time of 30\,d is observed above
20\,\kev, it is only 20\,d below 15\,\kev. Such behavior is typically
observed in systems where an irradiated disk is present which,
however, is not the case for \V. Since $L_\text{X} \propto
\dot M$, this picture suggests that $\dot M \propto M_\text{disk}$.
The transition to the linear phase would then be triggered by a yet
unknown change in the disk.

\section{Spectral evolution}

To study the evolution of the spectrum over the outburst, we used the
simple \textsf{cutoffpl} model, modified by two Gaussian absorption
lines to model the CRSFs at \ca27\,\kev and \ca51\,\kev for all
observations. While the folding energy remains constant at
\ca7.5\,\kev, the power law index $\Gamma$ decreases from $-0.18$ in
the first observation to $-0.4$ in the last observations -- the
spectrum of \V hardens over the outburst.  The fundamental cyclotron
line also changes over the outburst: the energy increases from
27.5\,\kev in the first \integral observation to 29.5\,\kev in the
last observations. Moreover, during the previous \xte observation, the
fundamental CRSF was observed at 26.3\,\kev \citep{pottschmidt05a}
resulting in a total increase of more than 3\,\kev.  This change is
highly significant: fitting the last \integral observations with a
CRSF energy fixed to 27.5\,\kev results in strong residuals and a
completely unacceptable fit.  The same holds true for the continuum:
fixing the other continuum parameters also results in unacceptable
fits. The determination of the parameters of the second CRSF, however,
is problematic for the second half of the observations as with
decreasing flux, statistics become poor.

\begin{figure}
\centerline{\includegraphics[width=0.9\columnwidth]{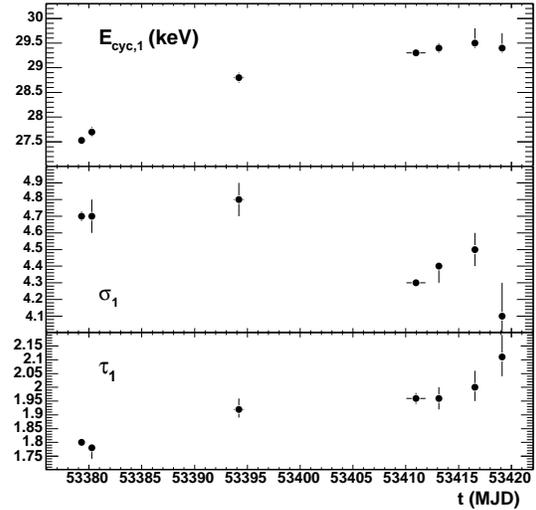}}
\caption{Evolution of the spectral parameters of the fundamental CRSF
  during the decay of the outburst \citep{mowlavi05a}.}
\label{fig:evol}
\end{figure}

\section{Discussion}

The exponential decay of the flux and the transition to a linear phase
later is frequently observed in SXTs and dwarf novae
\citep{king98a}. While the emission mechanism is entirely different
for \V, the similarity is striking and a yet unidentified change in
the disk can be assumed to trigger the transition to the linear phase.
The luminosity dependence of the energy of CRSFs had already been
observed previously \citep{mihara95a} and was assumed to be due a
change in height of the CRSF formation region in the accretion column.
Based on our data, we derive a change in height of \ca300\,m; however,
a slightly different picture is also possible: the CRSF emission
region can be assumed to be extended along the accretion column. The
observed broad CRSFs would then be superposition of many narrower
lines, each from a different height in the column. As the accretion
rate drops, the extend of the emission region and its height both
decrease and hence the energy of the CRSF increases while it gets
narrower as is observed for \V (see
Fig.~\ref{fig:evol}). 

\end{document}